# Initiatives Based on the Psychology of Scarcity Can Increase Covid-19 Vaccinations

Alessandro Del Ponte (corresponding author)

Assistant Professor, Department of Political Science, The University of Alabama

318 ten Hoor Hall, Marrs Spring Rd

Tuscaloosa, AL 35401, United States

Audrey De Dominicis

PhD Candidate, Department of Political Science, University of Teramo

via R. Balzarini 1

64100 Teramo, Italy

Paolo Canofari

Associate Professor, Department of Economic and Social Sciences, Marche Polytechnic University

Piazzale R. Martelli 8

60121 Ancona, Italy

**Abstract**

Background: Here we investigate whether releasing Covid-19 vaccines in limited quantities and at limited times boosted Italy's vaccination campaign in 2021. This strategy exploits insights from psychology and consumer marketing.

Methods: We built an original dataset covering 200 days of vaccination data in Italy, including "open day" events. Open-day events (in short: open days) are instances where Covid-19 vaccines were released in limited quantities and only for a specific day at a specified location (usually, a large pavilion or a public building). Our dependent variables are the number of total and first doses administered in proportion to the eligible population. Our key independent variable is the presence of open-day events in a given region on a specific day. We analyzed the data using regression with fixed effects for time and region. The analysis was robust to alternative model specifications.

Findings: We find that when an open day event was organized, in proportion to the eligible population, there was an average 0.39-0.44 percentage point increase in total doses administered and a 0.30-0.33 percentage point increase in first doses administered. These figures correspond to an average increase of 10,455-11,796 in total doses administered and 8,043-8,847 in the first doses administered.

Interpretation: Releasing vaccines in limited quantities and at limited times by organizing open-day events was associated with an increase in Covid-19 vaccinations in most Italian regions. These results call for wider adoption of vaccination strategies based on the limited release of vaccines for other infectious diseases or future pandemics.

**Introduction**

Vaccines against Covid-19 have been a key pillar of the public health strategy to mitigate the Covid-19 pandemic [1–3], and yet many citizens around the world have been reluctant to get vaccinated [4–6]. In response to the complex landscape of vaccine hesitancy, researchers and public authorities have attempted a broad range of communication initiatives and behavioral strategies to encourage citizens to receive the shot [7,8]. In Italy -the first epicenter of the Covid-19 outbreak outside of China- a key measure aimed at increasing vaccinations was the introduction of 'Open-Day' distribution initiatives [9]. In Italy's regionalized healthcare system, regional administrations oversaw "open day" events.

A vaccine open day means that a limited quantity of Covid-19 vaccines is released *only for that day* and at a particular location (typically, a large fair pavilion that accommodates large crowds). Amongst many different initiatives and strategies that have been rolled out around the world, open-day events in Italy are interesting from a public health perspective because they build on basic features of human psychology, such as the fear of missing out (FOMO) [10] and the psychology of scarcity [11,12]. Since they exploit basic traits of human nature, these strategies potentially lend themselves to tackling a variety of public health issues. Notably, leveraging natural scarcity or creating FOMO motivations were listed among the strategies for promoting Covid-19 vaccinations proposed by Wood and Schulman in the New England Journal of Medicine (NEJM) in 2021 [8]. The proposals in the NEJM rest on solid ground in the psychology of consumer behavior [12,13] but still await empirical validation. To satisfy this need, in this paper, we leverage a novel dataset on Covid-19 vaccinations in Italy to test the efficacy of open-day events as a strategy to increase the uptake of Covid-19 vaccines in the population.

Previous research in the field of marketing found that when consumers have the perception that a product is hardly available, they may attribute a higher value to that product. Thus, companies strategically create perceptions of product scarcity to increase consumers' interest in them. Typical strategies used in marketing include restricting the timing of sales, restricting the quantity released of a given product, or a combination of both strategies [12,13]. Creating a sense of scarcity is so vital to boosting sales that famous psychologist of influence Robert Cialdini dubbed scarcity as one of the six principles of persuasion [14]. Specifically, a meta-analysis of hundreds of effect sizes in studies of product marketing found that creating a sense that a product is scarce through limited release for a limited time works particularly well when consumption of the product is conspicuous and requires high involvement from the consumer [13]. Covid-19 vaccines share these features, to the point that people often posted vaccination selfies on social media and shared their post-vaccine experiences [15,16].

Together, insights from psychology and marketing inform the hypothesis that releasing Covid-19 vaccines in limited quantities and for a limited time (open-day events) should increase the administration of Covid-19 vaccines compared to the status quo where vaccines are released outside of these special events. Of course, this *limited-release hypothesis* takes into account the number of vaccine doses available and the number of patients eligible. To test the limited-release hypothesis, we turn to our original dataset from Italy.

## Methods

*Dataset*

We constructed the original dataset used in this paper starting from daily Covid-19 vaccination data in Italy ranging from December 27, 2020, to July 14, 2021, available on GitHub at the following link: https://github.com/italia/covid19-opendata-vaccini. From this dataset, we extracted the vaccine doses delivered to each of the 20 Italian regions; the number of first- and total Covid-19 vaccine doses administered daily; the region where the doses were administered to patients; and information about patient eligibility to get vaccinated. We augmented this dataset with data about open-day vaccination events held between April and July 2021 in each region. We collected information about open-day events from publicly available announcements reported by online local and regional Italian newspapers. In some cases, we were able to collect detailed information about each open-day event, including the city, the exact address (e.g., a hospital), the number of available doses, the brand of the vaccines available, and the age groups at which these initiatives were aimed. However, most of the time, the available data includes only whether a specific number of open-day events were held in a given region on a certain day. For this reason, the analysis is not partitioned by vaccine brand and does not include demographic information about vaccine recipients. As a result, our final dataset includes 200 data points (one per day) for each of the 20 Italian regions. Figure 1 shows the number of regions that organized at least one open-day event for every week in our sample.

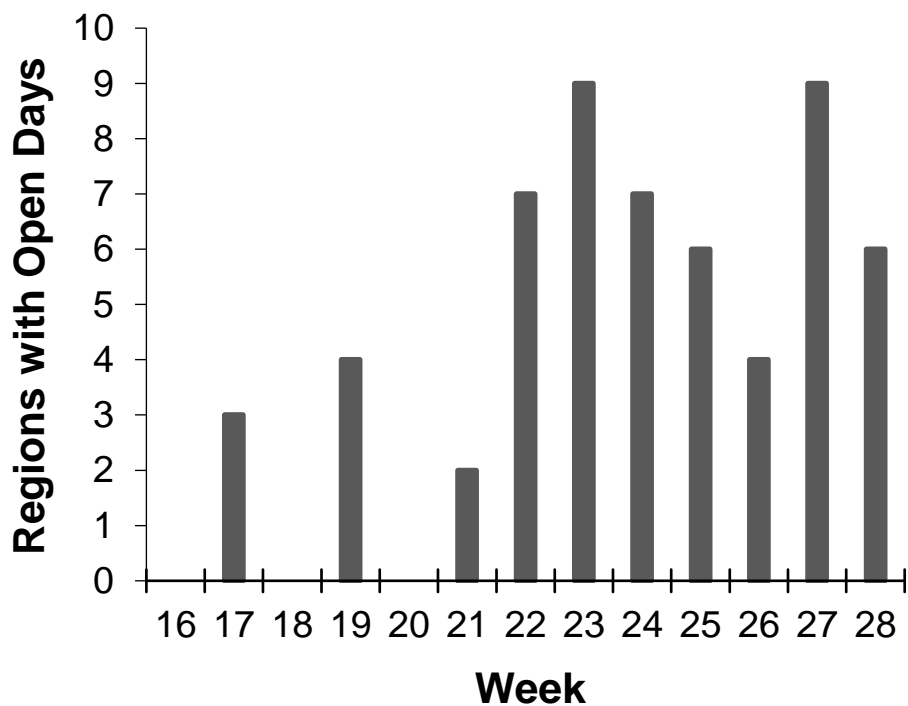

**Figure 1**. The number of regions organizing at least one open-day event from the third week of April 2021 to the second week of July 2021. The figure shows that the number of regions organizing at least one open-day event peaked in the last weeks of May and June.

*Statistical analysis*

To investigate the relationship between vaccine administrations and open-day events, we examined the period from December 27, 2020, when vaccines started to become available, until July 14, 2021, when the data collection ended. We constructed a dichotomous variable that is equal to 1 if at least one open-day event was organized that day, or 0 if no open-day event took place. In our analytical strategy, we study the impact of open day events on Covid-19 vaccinations by using as our main dependent variable the total doses administered/eligible population ratio, where the eligible population includes patients aged 12 or older.

Specifically, we estimate the following equation:

$$\frac{doses\ administered}{eligible\ population_{i,t}} = \beta_0 + \beta_1 \big(open\ day\ event_{i,t}\big) + \beta_2 Z_i + \varepsilon_{i,t} \quad [1]$$

where the dependent variable is the ratio between the number of administrations and the number of people eligible to receive the vaccine, $open\ day\ event_{i,t}$ is the open-day dummy variable, $Z_i$ represents unobserved time-invariant heterogeneities across regions, $\varepsilon_{i,t}$ is the error term, and $i\ and\ t$ are a region and time index, respectively. We use a panel regression with fixed effects for region and time, since statistically significant Hausman tests consistently showed that the fixed effects models outperformed panel regressions with random effects (a model comparison is available in Table A1). We carried out all our analyses using the Stata 14 statistical software. We report two-tailed tests of significance for all analyses. However, since the limited-release hypothesis is directional, we also report the results of one-tailed tests of significance.

**Results**

First, we report the descriptive statistics of the study variables. Table 1 shows the main descriptive statistics of the variables present in our dataset. It is worth mentioning a few notable figures. The number of daily doses administered averaged approximately 15,000, with a maximum of 117.252. The dataset includes a binary open-day variable, which keeps track of whether *any* number of open-day events were organized on a given date, and a count variable, which records the specific *number* of open-day events that were organized on a single day. The count variable ranges from 0 to 21 open-day events recorded. In some instances, there were caps to patients' eligibility to receive the vaccine. These caps reflected age targets for vaccine administrations. Minimum age eligibility was as low as 12 years old and as high as 80. Maximum age eligibility ranged from 16 to 100 years old. The number of doses delivered was typically positive but could also be occasionally negative when doses were returned. The number of eligible individuals ranged from just over 100,000 to just shy of nine million, which depends on the population size in a given region. Finally, we report descriptive statistics for the ratios between key measures and the eligible population.

**Table 1**. Descriptive statistics.

| Variable | N | M | SD | Min | Max |
|---|---|---|---|---|---|
| Doses administered (daily) | 3,916 | 15,012.74 | 18,892.53 | 1 | 117,252 |
| Open day (Yes/No) | 4,000 | 0.040 | 0.20 | 0 (No) | 1 (Yes) |
| Open day (Count) | 4,000 | 0.11 | 0.96 | 0 | 21 |
| Minimum age eligibility | 144 | 32.54 | 18.61 | 12 | 80 |
| Maximum age eligibility | 144 | 88.19 | 22.85 | 16 | 100 |
| Doses delivered | 2,059 | 30,917.45 | 64,327.88 | -49,955 | 779,713 |
| First doses administered | 3,916 | 9,005.31 | 12,029.30 | 0 | 93,189 |
| Eligible individuals | 4,000 | 2,680,845 | 2,191,825 | 112,534 | 8,975,783 |
| Doses delivered/eligible population ratio | 2,059 | 0.011 | 0.02 | -0.01 | 0.09 |
| First doses administered/eligible population ratio | 3,916 | 0.0032 | 0.0025 | 0 | 0.0139 |
| Total doses administered/eligible population ratio | 3,916 | 0.0055 | 0.0038 | 2.56e-07 | 0.02 |

*Note*. N = observations; M = mean; SD = standard deviation; Min = minimum; Max = maximum. A negative number of doses delivered means that some doses were returned before being administered to patients.

Next, we look at the results of our analysis that aims at testing the limited-release hypothesis. When at least one open-day event was present, the total doses administered relative to the eligible population increased by 0.44 percentage points (Table 2). In regions with large populations such as Campania and Tuscany, this effect translates into 22,359 and 14,709 additional doses administered, respectively. Meanwhile, in more sparsely populated regions such as Trentino Alto Adige and Marche, this effect corresponds to 4,188 and 6,012 additional doses, respectively. As a robustness check, in a second model, we control for the doses delivered relative to the eligible population (Table 2, model 2). Compared to the previous model, the coefficient for open-day events is slightly lower, but it always remains statistically positive and significant (0.39 percentage point increase in vaccines administered relative to the eligible population).

To further investigate the effect of open-day events on vaccinations, we estimate these models again, this time using the *number* of open-day events instead of the binary measure of open-day events. As shown in Table 1, the number of open-day events ranges from 0 to 21 across regions, allowing us to estimate the marginal effect of organizing an additional open-day event. These models yield similar results (Table A2). With each additional open-day event, the total doses administered relative to the eligible population increased by 0.05 percentage points. In highly populated regions such as Tuscany and Campania, this translates into 1,671 and 683 additional doses, respectively. In less populous regions such as Calabria and Marche, this effect corresponds to 850 and 683 additional doses, respectively.

**Table 2**. Regressions of the *total* doses administered/eligible population ratio on the binary open day variable and the doses delivered/eligible population ratio.

|  | Model 1 | | | Model 2 | | |
|---|---|---|---|---|---|---|
|  | **Coef** | **SE** | | **Coef** | **SE** | |
| **Open day (Yes/No)** | 0.0044 | 0.0003 | *** | 0.0039 | 0.0004 | *** |
| **Doses Delivered/Eligible population ratio** | _ | _ | | 0.0543 | 0.0048 | *** |
| **Constant** | 0.0053 | 0.00006 | *** | 0.0056 | 0.00009 | *** |

*Note*. Model 1: N = 3,916. Model 2 = N = 2,050. Coef. = Beta coefficient. SE = Standard error. The fixed effects model includes region and time fixed effects. * $p < 0.05$ ** $p < 0.01$ *** $p < 0.001$. The significant Hausman test ($p < .001$) indicates that the fixed effects model outperforms the random effects model.

The marginal effect of open-day events disaggregated by region is displayed in Figure 2. The left panel indicates the marginal effects of organizing at least one open day. The left panel reveals that the largest effects were in Sardinia, Trentino Alto Adige, and Campania. The open-day coefficients were positive and statistically significant in most regions (two-tailed tests). Exceptions include Basilicata, Calabria, Emilia Romagna, Liguria, Apulia, Valle d'Aosta, and Veneto, where the confidence interval overlaps with zero. However, considering that the hypothesized effect of open days is positive, if we use a one-tailed test, all coefficients are statistically significant except for Basilicata, Valle d'Aosta, and Veneto.

The right panel plots the marginal effects of open days using the count variable. Thus, it indicates the marginal effect of organizing an additional open day. The largest effects of an additional open day were in Lazio and Sardinia. Consistent with the binary measure, most coefficients were positive and statistically significant, with the same exceptions as before (two-tailed tests). If instead we consider the directional hypothesis of a positive effect (one-tailed tests), then coefficients are positive and significant in all regions but Basilicata and Veneto.

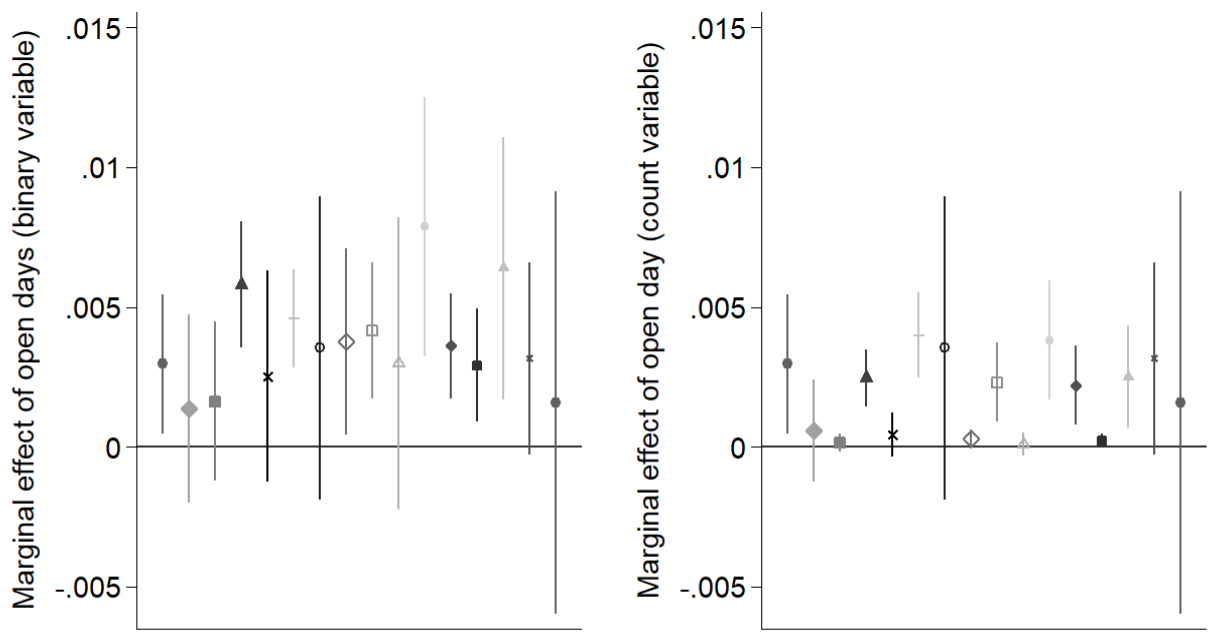

**Figure 2**. The marginal effect of organizing open day events in a specific region on *total* doses administered. The error bars indicate 95% confidence intervals. The panel on the left shows the marginal effects of open days using the binary measure (Yes/No). The panel on the right displays the marginal effects of open days using the count variable, and thus indicates the effect of organizing an additional open day. Four regions (Friuli Venezia Giulia, Lombardy, Molise, Umbria) are omitted because no open-day events were recorded there in the period considered in this study.

Next, we focus on the people who had never been vaccinated against Covid-19 (Table 3). The results show that, even in this case, the presence of open days is associated with a 0.33 percentage point increase in first dose administrations, which in a large region such as Tuscany corresponds to 11,032 additional first doses administered relative to the eligible population. Adding the control variable to this model, the results remain similar, with a 0.30 percentage point increase in the first doses administered. Taking again the example of Tuscany, this is equivalent to 10,029 extra first doses administered in proportion to the eligible population.

We repeat this analysis by examining the effect of each additional open-day event instead of the binary open-day event variable. With each additional open-day event, there was a 0.04 percentage point increase in first dose administrations (Table A3). In a large region such as Tuscany, this effect corresponds to 1,337 additional first doses administered relative to the eligible population. Adding the control variable to this model, the results remain similar, with a 0.03 percentage point increase in the first doses administered. Keeping Tuscany as an example, this effect is equivalent to 1,003 extra first doses administered in proportion to the eligible population.

**Table 3**. Regressions of the *first* doses administered/eligible population ratio on the binary open day variable and the doses delivered/eligible population ratio.

|  | Model 1 | | | Model 2 | | |
|---|---|---|---|---|---|---|
|  | Coef | SE | | Coef | SE | |
| **Open day (Yes/No)** | 0.0033 | 0.0002 | *** | 0.0030 | 0.0002 | *** |
| **Doses Delivered/Eligible population ratio** | _ | _ | | 0.0304 | 0.0032 | *** |
| **Constant** | 0.0032 | 0.00004 | *** | 0.0033 | 0.00006 | *** |

*Note*. Model 1: N = 3,916. Model 2 = N = 2,050. Coef. = Beta coefficient. SE = Standard error. The fixed effects model includes region and time fixed effects. * $p < 0.05$ ** $p < 0.01$ *** $p < 0.001$. The significant Hausman test ($p < .001$) indicates that the fixed effects model outperforms the random effects model.

We include a graphical presentation of the marginal effects of open days also for the analysis of the first doses administered in proportion to the eligible population (Figure 3). The left panel shows the effect of open days using the binary measure. Thus, it graphs the effect of organizing at least one open day in each region that organized them. We find that the largest effect was in Sardinia, Campania, and Trentino Alto Adige. Similar to total doses, the coefficient for open days was positive and significant in most regions, this time with fewer exceptions. In Abruzzo, Basilicata, Calabria, Liguria, Apulia, and Veneto, the confidence intervals overlap with zero, indicating no statistically significant effects when using two-tailed tests. However, if we use one-tailed tests following the directional hypothesis of a positive effect of open days, all coefficients are positive and statistically significant except for Calabria, where the point estimate is negative, although statistically indistinguishable from zero.

Looking at the marginal effects of organizing an additional open day (Figure 3, right panel), we find that the largest effects were in Valle d'Aosta and Sardinia. Again, coefficients were positive and significant in most regions, with even fewer exceptions: Abruzzo, Basilicata, Calabria, Liguria, Apulia, and Veneto (two-tailed tests). However, considering a directional hypothesis and using a one-tailed test, the coefficient for open days remains insignificant only in Calabria, where the point estimate is exactly zero.

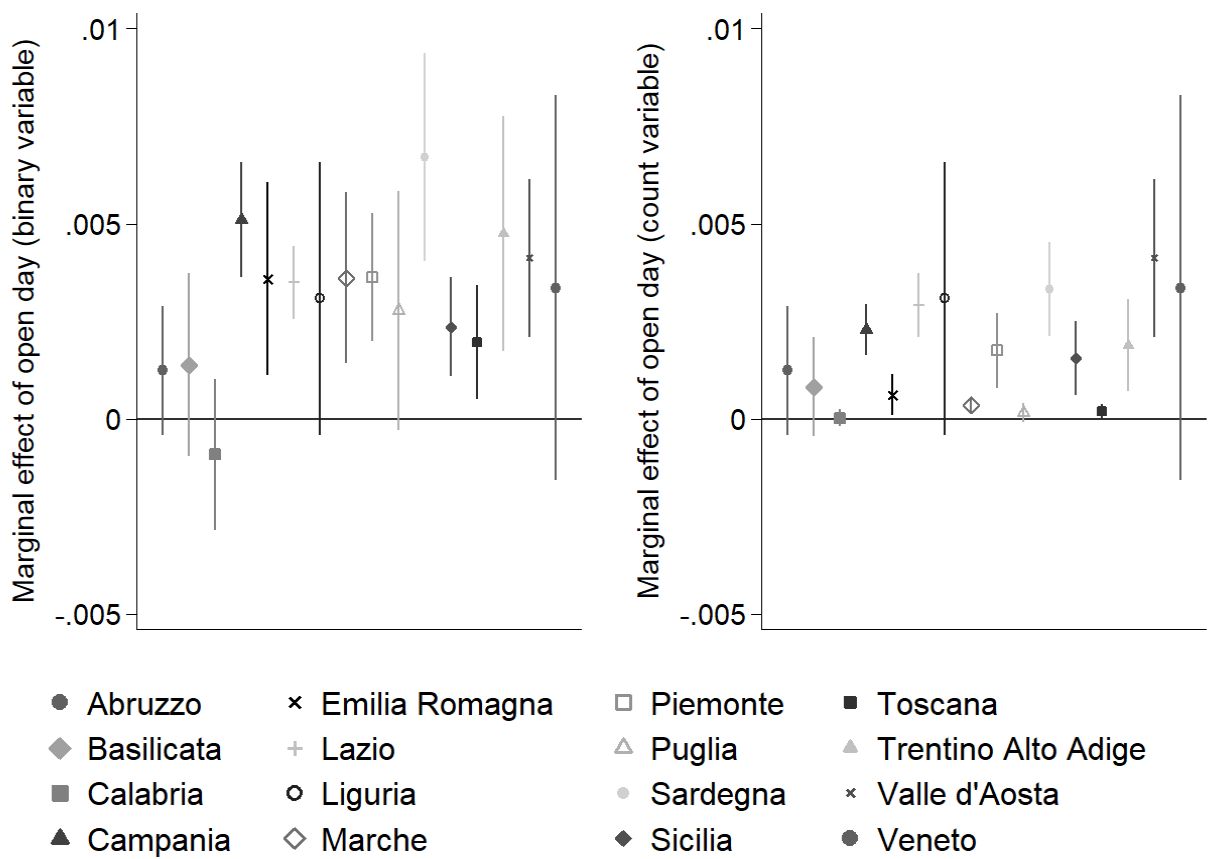

**Figure 3**. The marginal effect of organizing open day events in a specific region on the *first* doses administered. The error bars indicate 95% confidence intervals. The panel on the left shows the marginal effects of open days using the binary measure (Yes/No). The panel on the right displays the marginal effects of open days using the count variable, and thus indicates the effect of organizing an additional open day. Four regions (Friuli Venezia Giulia, Lombardy, Molise, Umbria) are omitted because no open-day events were recorded there in the period considered in this study.

**Discussion**

We tested the limited-release hypothesis, which predicts that releasing Covid-19 vaccines in limited quantities on select days ("open days") boosts the number of doses administered. Using an original dataset of open-day events organized in sixteen Italian regions between April and July 2021, we find support for the limited-release hypothesis: open-day events had a positive effect on vaccinations in most regions. Consistent with expectations, open days boosted administrations of both total and first doses. These effects were robust to different econometric specifications, including the use of fixed or random effects, controlling for doses delivered, and accounting for the population eligible to receive the vaccine. The results support the proposals by Wood and Schulman in the NEJM to exploit basic features of human psychology such as fear-of-missing-out (FOMO) motivations and the psychology of scarcity in order to increase the number of vaccinations against Covid-19 [8].

This research comes with natural limitations which future research could address. Importantly, our dataset covers only Italy and is purely observational because the organization of open-day events was not randomly assigned but was instead decided by each region. A possible concern is that regions that were more capable to organize open-day events self-selected themselves into offering these initiatives. However, this possibility seems unlikely since among the four regions which did not organize there is Lombardy, which is widely recognized as providing one of the best public healthcare systems in Italy [17]. More generally, the fact that 75% of Italian regions organized at least one open day mitigates worries about selection effects. Also, in addition to comparisons between regions, our panel dataset allows us to make within-region comparisons for weeks with and without open-day events, which further enhances the validity of the dataset to make inferences about the association between open-day events and an increase in Covid-19 vaccinations.

Another possible limitation associated with an observational dataset is that different regions may have organized open-day events differently. For example, some regions may have been more efficient; may have better communicated the rollout of these initiatives with the public; may have given additional small incentives for receiving the vaccinations; or, they may simply have managed these initiatives better overall. These factors may explain why the coefficient sizes differed across regions and why in Calabria -a region which consistently ranks at the bottom in Italy for the quality of healthcare [17], open day events did not significantly increase vaccinations in proportion to the eligible population. However, in our main analysis, the coefficient for open days was robust to the inclusion of region fixed effects which capture regional differences. Thus, the robustness of the results to region fixed effects allays concerns over different implementations of the open day events at the regional level.

Overall, these findings offer actionable insights for how to reach a wider population in future vaccination campaigns against coming pandemics. Future research could test the effectiveness of limited releases of vaccines in controlled settings, such as laboratory or field experiments, where researchers can reap the benefits of experimental control. To this end, testing the effects on the uptake of vaccines other than the ones against Covid-19 could also prove helpful. Since Covid-19 vaccines have become available only recently, it is plausible that the behavioral effects of limited vaccine releases might be different for vaccines that have been available for decades, such as influenza vaccines. For example, the effects could be larger because of less fear toward the vaccines or smaller because people may consider Covid-19 as a larger health threat than other viruses such as the flu, and thus be less susceptible to FOMO motivations and the psychology of scarcity. We hope that future work will explore these extensions, since vaccinating the largest possible patient population is critical to mitigating or even eradicating many infectious diseases which are a threat to global public health.

# Appendix

**Table A1**. Model comparison of the regressions of the *total* doses administered/eligible population ratio on the number of open day events and the doses delivered/eligible population ratio.

|  | Pooled OLS | | | Fixed Effects | | | Random Effects | | |
|---|---|---|---|---|---|---|---|---|---|
|  | Coef. | SE | | Coef. | SE | | Coef | SE | |
| **Open day (count)** | 0.00038 | 0.0001 | ** | 0.0004 | 0.0001 | *** | 0.0004 | 0.0001 | *** |
| **Doses delivered/eligible population ratio** | 0.0510 | 0.0029 | *** | 0.0554 | 0.0049 | *** | 0.0511 | 0.0049 | *** |
| **Constant** | 0.0058 | 0.0001 | *** | 0.0057 | 0.0001 | *** | 0.0058 | 0.0001 | *** |
| **Hausman Test** | | No | | | | | 0.0000 | | |

*Note*. N = 2,050. Coef. = Beta coefficient. SE = Standard error. The pooled OLS model includes standard errors clustered at the region level. The fixed effects model includes region fixed effects. * $p < 0.05$ ** $p < 0.01$ *** $p < 0.001$. The significant Hausman test indicates that the fixed effects model outperforms the random effects model.

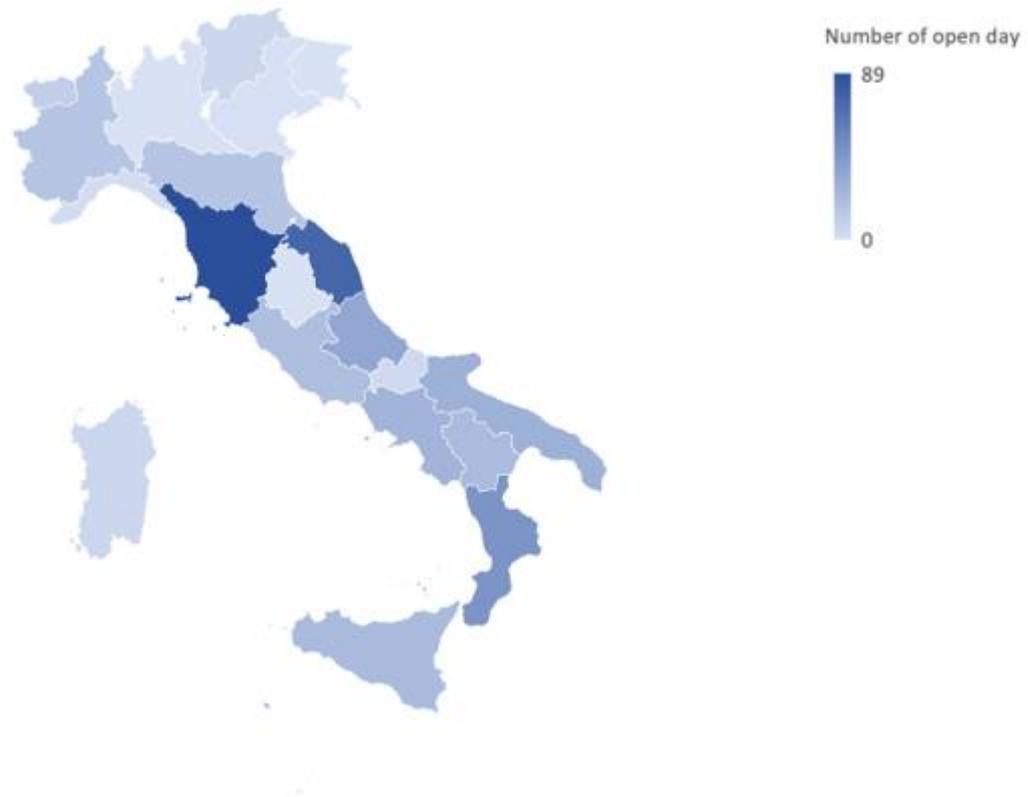

**Figure A1**. Number of open days in each region (from April 2021 to the second week of July 2021). The figure indicates that Central and Southern Italy organized more open vaccination events compared to the North, with Tuscany holding the highest number of open days (89).

**Table A2**. Regressions of the *total* doses administered/eligible population ratio on the number of open day events and the doses delivered/eligible population ratio.

|  | Model 1 | | | Model 2 | | |
|---|---|---|---|---|---|---|
|  | **Coef.** | **SE** | | **Coef.** | **SE** | |
| **Open day (count)** | 0.0005 | 0.0001 | *** | 0.0004 | 0.0001 | *** |
| **Doses Delivered/Eligible population ratio** | _ | _ |  | 0.0554 | 0.0049 | *** |
| **Constant** | 0.0055 | 0.0001 | *** | 0.0057 | 0.0001 | *** |

*Note*. Model 1: N = 3,916. Model 2 = N = 2,050. Coef. = Beta coefficient. SE = Standard error. * $p < 0.05$ ** $p < 0.01$ *** $p < 0.001$. Models include region- and time fixed effects. The significant Hausman test ($p < .001$) indicates that the fixed effects model outperforms a random effects model.

**Table A3**. Regressions of the *first* doses administered/eligible population ratio on the number of open day events and the doses delivered/eligible population ratio.

|  | Model 1 | | | Model 2 | | |
|---|---|---|---|---|---|---|
|  | Coef. | SE | | Coef. | SE | |
| **Open day (count)** | 0.0004 | 0.00004 | *** | 0.0003 | 0.00005 | *** |
| **Doses Delivered/Eligible population ratio** | _ | _ | | 0.0312 | 0.0033 | *** |
| **Constant** | 0.0032 | 0.00004 | *** | 0.0034 | 0.00006 | *** |

*Note*. Model 1: N = 3,916. Model 2 = N = 2,050. Coef. = Beta coefficient. SE = Standard error. * $p < 0.05$ ** $p < 0.01$ *** $p < 0.001$. Models include region- and time fixed effects. The significant Hausman test ($p < .001$) indicates that the fixed effects model outperforms a random effects model.